\documentstyle[12pt]{article}
\title{Families of exact solutions of a 2D gravity model minimally coupled to electrodynamics}
\author{S. K. Moayedi$^{a}$,
F. Darabi$^{b, c}$\thanks{Corresponding author, e-mail: f-darabi@cc.sbu.ac.ir}\\
$^{a}${\small Department of Physics, Arak University, Arak, Iran.}\\
$^{b}${\small Department of Physics, Tarbiyat Moallem University of Tabriz, Tabriz, Iran.}\\
$^{c}${\small Department of Physics, University of Waterloo, Ontario, N2L 3G1, Canada.}
}
\begin{document}
\maketitle
\baselineskip.4in
\begin{abstract}
Three families of exact solutions for 2-dimensional gravity minimally coupled to
electrodynamics are obtained in the context of ${\cal R}=T$ theory.
It is shown, by supersymmetric formalism of quantum mechanics, that the quantum
dynamics of a neutral bosonic particle
on static backgrounds with both varying curvature and electric field is exactly solvable.
\end{abstract}
\vspace{2cm}
\section*{I Introduction}

It is well known that Einstein's gravitational theory in 2-dimensions is trivial.
This is because the Einstein tensor is identically zero for all 2-dimensional metrics.
Consequently, the Einstein equations require the energy-momentum tensor to be vanished and
this result is inconsistent with some non-trivial matter configurations \cite{1}.
Also, in the mathematical language the Euler class $\chi$ of a compact and 2-dimensional
manifold $M$ with boundary $\partial M$ is defined by \cite{2}
$$
2 \pi \chi=\frac{1}{2} \int_{M} d^2 x \sqrt{-g}{\cal R}-\int_{\partial M} \sqrt{-g} K,
$$
where ${\cal R}$ is the Ricci scalar and $K$ is the trace of the extrinsic curvature $K_{ij}$.
Therefore, if we demand more useful information on a 2-dimensional space-time a different gravitational action
is required. An interesting action for 2-dimensional gravity has been derived in the context of string theory \cite{3}.
However, despite the string theory approach it has been recently shown that gravity in 2-dimensions is not necessarily
trivial \cite{1, 4}. Of particular interest among these new approaches is the ${\cal R}=T$ theory of 2-dimensional
gravity in which the scalar curvature ${\cal R}$ is equal to the trace of the energy-momentum tensor $T$ \cite{1, 4}.
This theory has provided some remarkable classical and semi-classical results such as: well defined Newtonian limit
\cite{5}, black hole solutions \cite{4, 6}, gravitational radiation, FRW cosmology, gravitational collapse \cite{7},
black hole radiation \cite{8} and their thermodynamical properties \cite{4}. On the other hand, there are some similarities
between 4-dimensional and 2-dimensional gravity. Therefore, the study of classical and quantum behavior of ${\cal R}=T$ theory of
2-dimensional gravity may help us to get a deeper insight into the problems involved in 4-dimensional gravity.

In this paper\footnote{We use the units in which $\hbar=c=8 \pi G=1$.},
we find families of exact solutions of 2-dimensional
${\cal R}=T$ type gravity minimally coupled to electrodynamics.
Then, we investigate the quantum dynamics of a neutral bosonic particle on the obtained class
of static solutions by using the supersymmetric
quantum mechanics and obtain the energy spectrum and eigenfunctions exactly.
In section II and III, we introduce the model and obtain three families
of exact solutions for : a) 2-dimensional manifolds with constant
curvature and varying electric field, b) 2-dimensional manifolds with varying curvature and constant electric field, and c)
2-dimensional manifolds with both varying curvature and electric field.
Then, in section IV the quantum dynamics of a neutral bosonic
particle on the obtained static backgrounds with both varying curvature and electric field
is investigated in the context of supersymmetric quantum mechanics.
The paper ends with a brief conclusion.

\section*{II The model}

It is well known that in two dimensions one can write locally the metric
in the form \cite{0}
\begin{equation}
g(X)=e^{\phi(X)} \left(\begin{array}{cc} 1 & 0 \\ \\ 0 & -1 \end{array}\right)
\label{1}
\end{equation}
where $X:=(t, x)$ and $\phi(X)$ is a scalar field. The Ricci scalar for the metric (\ref{1}) is given by
\begin{equation}
{\cal R}(X)=e^{-\phi(X)} (\partial^2_x -\partial^2_t) \phi(X).
\label{2}
\end{equation}
Now, we put the matter as a Maxwell field with strength tensor $F_{\mu \nu}(\mu, \nu=0, 1)$ on the above two dimensional
Lorentzian geometry. We are interested in ${\cal R}=T$ theory of 2-dimensional gravity in which ``$T$'' is the
trace of electromagnetic energy-momentum tensor given by
\begin{equation}
T_{\alpha \beta}=-\epsilon(F_{\alpha \mu} F_{\beta}^{\mu}-\frac{1}{4} g_{\alpha
\beta}F_{\lambda \sigma} F^{\lambda \sigma})
\label{3}
\end{equation}
where $|\epsilon|=1$ and its sign is arbitrary. The $T_{\alpha \beta}$ in Eq (\ref{3}) is the special case of one
which was already considered by Mann et. al. \cite{6} for charged point particles.

The coupled Einstein-Maxwell field equations are as follows \cite{6}
\begin{equation}
{\cal R}=T
\label{4}
\end{equation}
\begin{equation}
\frac{1}{\sqrt{-g}} \partial_{\nu} (\sqrt{-g} F^{\mu \nu})=J^{\mu}
\label{5}
\end{equation}
where the first is the Einstein gravity in 2-dimensions and the second is the Maxwell equation in 2-dimensional curved space-time
from which the current conservation is easily derived. We note that the electromagnetic field $F_{\mu \nu}$ in 2-dimensions
has only one independent non-zero component as electric field namely
\begin{equation}
F_{tx}=E(X).
\label{6}
\end{equation}
Therefore, considering the form of strength tensor $F_{\mu \nu}$ in Eq (\ref{6}) the trace of energy-momentum tensor (\ref{3}) is obtained as
\begin{equation}
T=\epsilon e^{-2 \phi} E^2.
\label{7}
\end{equation}
Substituting the scalar curvature (\ref{2}) and the trace (\ref{7}) into equation (\ref{4}) we have
\begin{equation}
(\partial_t^2 -\partial_x^2)\phi(X) + \epsilon e^{-\phi(X)} E^2(X)=0.
\label{8}
\end{equation}

\section*{III Three families of exact solutions}

Now, we classify the solutions of Einstein-Maxwell field equations in three categories:
\\
\\
{\bf a)  2D manifolds with constant scalar curvature and varying electric field}
\\
\\
Assuming the constant scalar curvature ${\cal R}(X)={\cal R}_0$ and considering equation (\ref{4}) we find the varying electric field as
\begin{equation}
E(X)=\pm \sqrt{\frac{{\cal R}_0}{\epsilon}} e^{\phi(X)}.
\label{9}
\end{equation}
In equation (\ref{9}) the parameter $\epsilon$ is suitably chosen so that
for negative or positive scalar curvature ${\cal R}_0$
the electric field is always a real quantity. In this way, for
${\cal R}_0<0$ (one-sheet hyperboloid or a half plane) and for ${\cal R}_0>0$ (a
strip or a half plane) we should correspond $T<0$ and $T>0$,
respectively. Using (\ref{9}), equation (\ref{8}) becomes \begin{equation}
(\partial_t^2-\partial_x^2) \phi(X)+{\cal R}_0 e^{\phi(X)}=0
\label{10}
\end{equation}
which is known as Liouville equation \cite{9}. It is worth noting that equation (\ref{10}) was considered, at first, in the study of
{\em pure mathematical theory} of 2-dimensional surfaces with constant curvature and
is derived here through the {\em physical} ${\cal R}=T$
theory of gravity in 2-dimensions with constant curvature. Of course,
it is not so surprising because in the matter coupled gravity theory
here, with dynamical variables $g_{\mu \nu}, F_{\mu \nu}$ and $\phi$, the condition
${\cal R}=T$ may give rise to a dynamical equation in
gravity (geometry) sector with variables ${\cal R}, \phi$, and by taking a constant
value for ${\cal R}$  it falls into the Liouville's mathematical context.
In other words, equation (\ref{10}) is the Liouville equation in which the
constancy of curvature is introduced through the physical theory as
${\cal R}=T=\mbox{Const.}$\\
General solutions of equation (\ref{10}) is given by \cite{10}
\begin{equation}
\phi(x^+, x^-)=\log \frac{8 A'_+(x^+) A'_-(x^-)}{|{\cal R}_0|[A_+(x^+)-\frac{|{\cal R}_0|}{{\cal R}_0}A_-(x^-)]^2}
\label{11}
\end{equation}
where
$$
A'_{\pm}:=\frac{dA_{\pm}}{dx^{\pm}}\:\:\:, \:\:\:x^{\pm}=t\pm x.
$$
Substituting the solutions for $\phi$, eq (\ref{11}), into equations (\ref{1}) and
(\ref{9}) the metric solution and electric field are explicitly obtained with
respect to arbitrary functions $A_{\pm}(x^{\pm})$ as
\begin{equation}
ds^2=\frac{8 A'_+(x^+) A'_-(x^-)}{|{\cal R}_0|[A_+(x^+)-\frac{|{\cal R}_0|}{{\cal R}_0}A_-(x^-)]^2} dx^+ dx^-
\label{12}
\end{equation}
\begin{equation}
E(x^+, x^-)=\pm \sqrt{\frac{{\cal R}_0}{\epsilon}}
\frac{8 A'_+(x^+) A'_-(x^-)}{|{\cal R}_0|[A_+(x^+)-\frac{|{\cal R}_0|}{{\cal R}_0}A_-(x^-)]^2}.
\label{13}
\end{equation}
Using equation (\ref{5}) one can easily show that the covariant current $J^{\mu}$ corresponding to the solutions (\ref{12}),
(\ref{13}) vanishes. The metric (\ref{12}) describes a family of 2-dimensional Lorentzian manifolds with the same constant curvature ${\cal R}_0$.
In \cite{10} it was shown that all 2-dimensional Lorentzian manifolds with
the same constant curvature are locally isometric.
Therefore, as a result, it may be said that all metrics defined by (\ref{12}) are locally isometric.
\\
\\
{\bf b) 2D manifolds with varying scalar curvature and constant electric field}
\\
\\
In this case, by assumption of a constant electric field as $E(X)=E_0$ equation (\ref{8}) is written as
\begin{equation}
(\partial^2_t -\partial^2_x)\phi(X)+\epsilon E_0^2 e^{-\phi(X)}=0.
\label{14}
\end{equation}
As before, one can show that general solution to equation (\ref{14}) is given by
\begin{equation}
\phi(x^+, x^-)=-\log \frac{8 B'_+(x^+) B'_-(x^-)}{E^2_0[B_+(x^+)+\frac{1}{\epsilon}B_-(x^-)]^2}
\label{15}
\end{equation}
where $B_{\pm}(x^{\pm})$ are arbitrary functions of their arguments. Considering the solution (\ref{15}) the metric solution
and the scalar curvature are given as
\begin{equation}
ds^2=\frac{E^2_0[B_+(x^+)+\frac{1}{\epsilon}B_-(x^-)]^2}{8 B'_+(x^+) B'_-(x^-)}dx^+ dx^-
\label{16}
\end{equation}
\begin{equation}
{\cal R}(x^+, x^-)=\frac{64
\epsilon}{E_0^2}\left[\frac{B'_+(x^+)B'_-(x^-)}{[B_+(x^+)+
\frac{1}{\epsilon}B_-(x^-)]^2}\right]^2.
\label{17}
\end{equation}
The current ${\bf J}$ corresponding to the solutions (\ref{16}) and
(\ref{17}) has the components
\begin{equation}
J^{\pm}=\mp 2E_0 e^{-2 \phi}\partial_{\mp}\phi
\label{18}
\end{equation}
where the functions $\phi$ are given by equation (\ref{15}).
\\
\\
{\bf c) 2D manifolds with varying scalar curvature and varying electric field}
\\
\\
In this case considering equations (\ref{2}) and (\ref{8}) the following relation between the scalar curvature and the electric field is obtained
\begin{equation}
E(X)=e^{\phi(X)} \sqrt{\frac{{\cal R}(X)}{\epsilon}}.
\label{19}
\end{equation}
For each given function $\phi(X)$ one can find, using equations (\ref{1}), (\ref{2}) and (\ref{19}), the corresponding metric solution and electric field.
Moreover, the current which produces the electric field (\ref{19}) is obtainable by equation (\ref{5}). For example, for $\phi$ as merely a function of spatial coordinate
$x$, namely for a static space-time, we find $J^x=0$ and that the static electric field is produced by $J^t$ component.
The importance of 2-dimensional manifolds with both varying scalar curvature and electric field in the static case, as will be shown in the
next section, is that the quantum dynamics of a neutral bosonic particle on
these manifolds is solvable exactly by using the
{\em Supersymmetric quantum mechanics} and {\em Shape invariance}.

\section*{IV Quantum dynamics of a neutral bosonic particle on static 2D
manifolds with varying scalar curvature and varying electric field }

Quantization of particle dynamics in 2-dimensions with constant
curvature for both massive and massless particles was investigated
in \cite{10}. Here, our aim is to quantize the particle dynamics
on the manifold discussed in part ``c'' of the previous section in the static case. Quantum dynamics of a massive neutral
bosonic particle is described by 2-dimensional Klein-Gordon equation
$$
\frac{1}{\sqrt{-g}}\partial_{\alpha}(\sqrt{-g} g^{\alpha \beta} \partial_{\beta})\Psi(X)+m^2\Psi(X)=0
$$
where $m$ is the mass of the particle. Assuming the scalar wave function $\Psi(X)$
as
$$\Psi(x, t)=e^{-i{\cal E}t} \psi(x)
$$
the Klein-Gordon equation on the 2D manifold obtained by the metric (\ref{1}) in the
static case ($\phi=\phi(x)$) becomes
\begin{equation}
\left[-\frac{d^2}{dx^2}+m^2(e^{\phi(x)}-1)\right]\psi(x)=({\cal E}^2-m^2)\psi(x).
\label{20}
\end{equation}
Equation (\ref{20}) is mathematically equivalent to one dimensional time
independent Schr\"{o}dinger equation. We use the supersymmetric quantum mechanics
to solve this equation. In supersymmetric quantum mechanics the {\em creation} and
{\em annihilation} operators are defined respectively as \cite{11, 12}
$$
{\cal A}^\dagger :=-\frac{d}{dx}+W(x)
$$
$$
{\cal A} :=\frac{d}{dx}+W(x)
$$
where $W(x)$ is called the {\em Superpotential}.
The supersymmetric {\em Partner} Hamiltonians namely $H_+={\cal A} {\cal A}^\dagger, H_-={\cal A}^\dagger {\cal A}$ have the following form
\begin{equation}
H_{\pm}=-\frac{d^2}{dx^2}+V_{\pm}(x)
\label{21}
\end{equation}
where the partner potentials $V_{\pm}(x)$ are given with respect to the superpotential $W(x)$ as
\begin{equation}
V_{\pm}(x)=W^2(x)\pm \frac{dW(x)}{dx}.
\label{22}
\end{equation}
If the partner potentials $V_{\pm}(a_0, x)$ (with $a_0$ as a constant parameter) are related according to the relation
\begin{equation}
V_+(a_0, x)=V_-(a_1, x)+R(a_1)
\label{23}
\end{equation}
then they are called {\em Shape-invariant} potentials \cite{13}. In equation (\ref{23}),
$a_1=F(a_0)$ is a new set of parameters and the term $R(a_1)$ is $x$ independent.
For shape-invariant potentials given by eq. (\ref{23}) the spectrum and
eigenfunctions are obtained by algebraic approach \cite{11}.
Comparing the left hand side of equation (\ref{20}) with equation (\ref{21}) we
deduce
\begin{equation}
m^2(e^{\phi_{\pm}(x)}-1)=V_{\pm}(x)
\label{24}
\end{equation}
which relates the conformal factors $e^{\phi_{\pm}(x)}$ to the partner potentials $V_{\pm}(x)$.
In fact, the supersymmetry and conformal degree of freedom let us to have two sets of 2-dimensional
static manifolds with varying curvature
$$
ds^2_{\pm}=e^{\phi_{\pm}(x)} (dt^2-dx^2)
$$
together with static electric fields $E_{\pm}(x).$ For example, by a suitable choice for the conformal factor  as
\begin{equation}
e^{\phi_-(x)}=\frac{\omega}{2m^2}(\frac{\omega}{2}x^2-1)+1
\label{25}
\end{equation}
we may consider the following superpotential
\begin{equation}
W(x)=\frac{1}{2}\omega x\:\:\:,\:\:\: \omega>0
\label{26}
\end{equation}
where $\omega$ is a quantity with the dimension of $(mass)^2$ or $(length)^{-2}$
in the units $\hbar=c=1$.
Now, in order to study the one set of solutions we calculate the partner potential
$V_-(x)$, by using equation (\ref{24}), as
\begin{equation}
V_-(x)=\frac{\omega}{2}(\frac{\omega}{2}x^2-1).
\label{27}
\end{equation}
Using equation (\ref{24}) we can obtain the energy and wave function
\cite{11, 12}
$$
{\cal E}^2_n=m^2+n\omega\:\:\:,\:\:\: n=0, 1, 2,...
$$
\begin{equation}
\psi_n(x)=C_n \exp(-\frac{\omega}{4}x^2) H_n(\sqrt{\frac{\omega}{2}}x)
\label{28}
\end{equation}
where $C_n$ is the normalization constant.
The static metric solution and electric field corresponding to the conformal factor
given by (\ref{25}) are as follows
\begin{equation}
ds^2_-=\left[1+\frac{\omega}{2m^2}(\frac{\omega}{2}x^2-1)\right](dt^2-dx^2)
\label{29}
\end{equation}
\begin{equation}
E_-(x)=\sqrt{\frac{\omega^2}{2m^2\epsilon}
\frac{1-\frac{\omega}{2m^2}(\frac{\omega}{2}x^2+1)}{1+\frac{\omega}{2m^2}
(\frac{\omega}{2}x^2-1)}}.
\label{30}
\end{equation}
The current corresponding to the electric field (\ref{30}) has the
non-vanishing
component
\begin{equation}
J^t(x)=-e^{-\phi_-(x)}\partial_x[e^{-\phi_-(x)}E_-(x)].
\label{31}
\end{equation}
It is easy to show that for $1-\frac{2 m^2}{\omega}\geq 0$ the metric (\ref{29}) is degenerate
at
\begin{equation}
\label{32}
x=\pm \sqrt{\frac{2}{\omega}(1-\frac{2m^2}{\omega})}.
\end{equation}
On the other hand, by calculating the Ricci scalar corresponding to the metric (\ref{29})
as
\begin{equation}
\label{33}
{\cal R}=\frac{\omega^2}{2m^2}\frac{1-\frac{\omega}{2m^2}-\frac{\omega^2x^2}{4m^2}}{[1+\frac{\omega}{2m^2}(\frac{\omega}{2}x^2-1)]^3}
\end{equation}
we find that the geometry defined by (\ref{29}) has also essential singularities at the
points (\ref{32}). The general existence of these {\em naked} singular points
indicates that the particle dynamics is exactly solvable
in the part of the manifold not including these singular points.
Alternatively, it seems possible to avoid the singular behavior only in a
sub-class of manifold by appropriate choices of the values on $\omega$ and $m$.
To this end, we may restrict ourselves to the values of $m$ and $\omega$ satisfying
the relation $1-\frac{2m^2}{\omega}<0$, which makes the metric (\ref{29}) free
of singularity. Then, we have
$$
\left\{\begin{array}{l}
{\cal R}=0 \quad \mbox{at the points}\quad x_0^{\pm}=\pm \sqrt{-\frac{2}{\omega}(1-\frac{2m^2}{\omega})},\\
{\cal R}>0 \quad \mbox{for the range}\quad x_0^-<x<x_0^+,\\
{\cal R}<0 \quad \mbox{for} \quad x>x_0^+ \quad\mbox{and}\quad x<x_0^-. \end{array}\right.
$$
Finally, we point out that for each partner potential $V_+(x)$ corresponding to the conformal factor
$e^{\phi_+(x)}$ we may obtain a set of static 2-dimensional manifolds with varying
curvature together with non-vanishing electric field with non-trivial current
distribution. Obviously, the same procedure may be exactly applied for other
shape-invariant potentials given by \cite{11, 12}.

\section*{Concluding remarks}

In this paper we have found three families of exact solutions for 2-dimensional
${\cal R}=T$
theory of gravity minimally coupled to electrodynamics.  By the study of
quantum dynamics of a neutral bosonic particle on a static 2-dimensional
space-time background
we have shown that: {\em Supersymmetric formalism of quantum mechanics leads to two
disjoint sets of static 2-dimensional manifolds with both varying curvature and electric field}. It is
possible to solve exactly the quantum dynamics of a neutral bosonic
particle
on these manifolds. It seems that the study of quantum dynamics of a fermionic
particle in 2-dimensional space-time may lead to 2-dimensional Lorentzian
manifolds as the solutions of 2D gravity coupled to electrodynamics and also to the quantum solvability of
particle dynamics on these manifolds \cite{14}.

\section*{Acknowledgment}

F. Darabi would like to thank the Department of Physics at the University of Waterloo
for their hospitality during his visit.


\newpage
\begin{thebibliography}{99}
\bibitem{1}R. B. Mann, in {\em general Relativity and Relativistic Astrophysics},
Proceedings of the 4th Canadian Conference, 1-15, (World Scientific, Singapore,
1992).
\bibitem{2}B. Jensen. J. Math. Phys. {\bf 38}, 1329 (1997); S. Chern, Ann. Math. {\bf 45}, 747 (1944); {\bf 46}, 674 (1945).
\bibitem{3}M. B. Green, J. H. Schwarz, E. Witten, Superstring Theory, (Cambridge
University Press, Cambridge, 1987).
\bibitem{4}R. B. Mann, ``Lower Dimensional Black Holes: Inside and Out'',
gr-qc/9501038
\bibitem{5}R. B. Mann, Found.Phys. Lett. 4, 425 (1991).
\bibitem{6}R. B. Mann, A. Shiekh, and L. Tarasov, Nucl. Phys. B341, 134 (1990).
\bibitem{7}A. E. Sikkema and R. B. Mann, Class. Quantum Grav. 8, 219 (1991).
\bibitem{8}R. B. Mann, S. Morsink, A. E. Sikkema and T. G. Steele, Phys. Rev. D43,
3948 (1991).
\bibitem{0}J. A. Wolf, {\em Spaces of constant curvature} (Boston: Publish
or Perish Inc 1974).
\bibitem{9}J. Liouville, J. Math. Pures Appl. 18, 71 (1853).
\bibitem{10}G. Jorjadze and W. Piechocki, Phys. Lett. B448, 203 (1999);
 ``Geometry of 2D spacetime and quantization of particle dynamics'', gr-qc/9811094.
\bibitem{11}G. Junker, ``{\em Supersymmetric Methods in Quantum and Statistical
Physics}'', (Springer-Verlag, 1996).
\bibitem{12}F. Cooper, A. Khare and U. Sukhatme, Phys. Rep. 251, 267 (1995).
\bibitem{13}L. E. Gendenshtein, JETP Lett. 38, 356 (1983).
\bibitem{14}Works are in progress.
\end{thebibliography}
\end{document}